\documentclass{PoS}

\usepackage{verbatim}
\usepackage{setspace}

\title{The web-PLOP observation prioritisation system\thanks{http://www.artemis-uk.org/web-PLOP}}

\ShortTitle{The web-PLOP observation prioritisation system}

\author{\speaker{Colin Snodgrass}\\
        European Southern Observatory, Alonso de Cordova 3107, Vitacura, Santiago, Chile\\
        E-mail: \email{csnodgra@eso.org}}

\author{Yiannis Tsapras, Rachel Street\\
        Las Cumbres Observatory, 6740 Cortona Dr. Suite 102, Santa Barbara, CA, USA\\
        E-mail: \email{ytsapras@lcogt.net}, \email{rstreet@lcogt.net}}

\author{Daniel Bramich\\
        Isaac Newton Group of Telescopes, Apartado de Correos 321, E-38700 Santa Cruz de la Palma, Canary Islands, Spain\\
        E-mail: \email{dmb@ing.iac.es}}

\author{Keith Horne, Martin Dominik\\
        Scottish Universities Physics Alliance, School of Physics and Astronomy, University of St Andrews, North Haugh, St. Andrews KY16 9SS, United Kingdom\\
        E-mail: \email{kdh1@st-and.ac.uk}, \email{md35@st-and.ac.uk}}

\author{Alasdair Allan\\
        School of Physics, University of Exeter, Stocker Road, Exeter EX4 4QL, United Kingdom\\
        E-mail: \email{aa@astro.ex.ac.uk}}

\abstract{We present a description of the automated system used by RoboNet to prioritise follow up observations of microlensing events to search for planets. The system keeps an up-to-date record of all public data from OGLE and MOA together with any existing RoboNet data and produces new PSPL fits whenever new data arrives. It then uses these fits to predict the current or future magnitudes of events, and selects those to observe which will maximise the probability of detecting planets for a given telescope and observing time. The system drives the RoboNet telescopes automatically based on these priorities, but it is also designed to be used interactively by human observers. The prioritisation options, such as telescope/instrument parameters, observing conditions and available time can all be controlled via a web-form, and the output target list can also be customised and sorted to show the parameters that the user desires.}

\FullConference{The Manchester Microlensing Conference: The 12th International Conference and ANGLES Microlensing Workshop\\
		 January 21-25 2008\\
		 Manchester, UK}

\begin{document}

\section{Introduction}

The Planet Lens OPtimisation (PLOP or web-PLOP)  tool was developed with two motivations; to provide an optimal target list for the automated observing of the RoboNet project \cite{Burgdorf07,Tsapras08}, and also to provide such lists to human observers at any telescope. It can be found at {\tt http://www.artemis-uk.org/web-PLOP/}. It is formed of two parts: 1.~A background code that keeps track of the current data on each event (from OGLE, MOA and RoboNet), and produces a new Point-Source-Point-Lens (PSPL) fit whenever new data arrives; 2.~A web form that gets parameters describing the telescope, observing conditions and total available observing time from the user. The results from the PSPL fits give the event parameters $t_{\rm E}$, $A_0$ etc.~that are used to predict the magnification and therefore brightness at the requested time of observation. Together with the parameters from the web form this allows prediction of the accuracy of the photometry that can be achieved on each event and calculation of the `worth' of observing it; by selecting the most worthwhile events we produce an optimal list of targets for the requested telescope at the requested time, with suggested exposure times. We describe our prioritisation algorithms in detail in \cite{keith}, but essentially we attempt to take the observations which will maximise the potential for discovery of a (planetary) anomaly, by maximising the `detection zone' area. The target list is outputted in either a machine readable or sortable human friendly format. Within RoboNet, this output dictates the target list that the eSTAR system \cite{estar} takes observations on, and new data is fed back into the PSPL model to close the loop and give priorities that are based on data just taken. For human observers at other sites, the output pages are customisable to display any desired parameter along with the priority of each event, and also show light-curves and detection zone maps along with links to the finding charts and original OGLE and/or MOA pages for each. Although written for RoboNet, this prioritisation tool is freely available and other observers are encouraged to make use of it.

\begin{figure}[htbp] 
   \centering
   \includegraphics[width=0.9\textwidth]{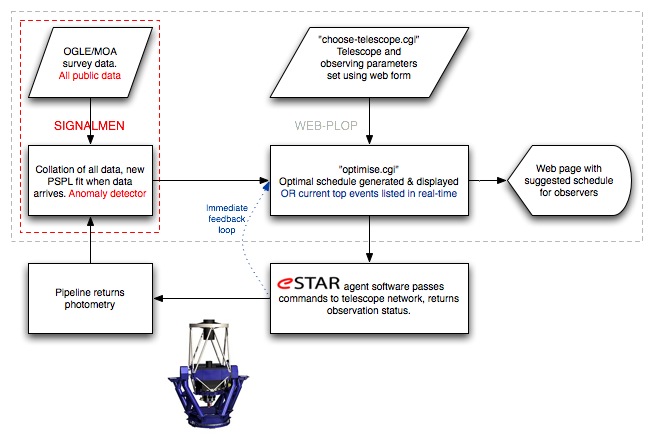} 
   \caption{The web-PLOP system and its interaction with RoboNet through eSTAR. The red box shows the elements that have overlap with SIGNALMEN, and the extra features that this system will provide in 2008.}
   \label{plop}
\end{figure}

Figure \ref{plop} shows an outline of what PLOP does, in terms of its interaction with the RoboNet system (through eSTAR). Starting with the 2008 Bulge season this system will 
provide not just an optimal schedule for a night's observing, but also real-time priorities that respond to the data taken by the RoboNet telescopes and alerts issued by the 
SIGNALMEN anomaly detector \cite{signalmen}. These new components are shown in blue and red on fig.~\ref{plop}, and are described in section \ref{realtime}. 

\section{Input Parameters}

\begin{figure} 
   \centering
\begin{tabular}{c c}
   \includegraphics[width=0.5\textwidth]{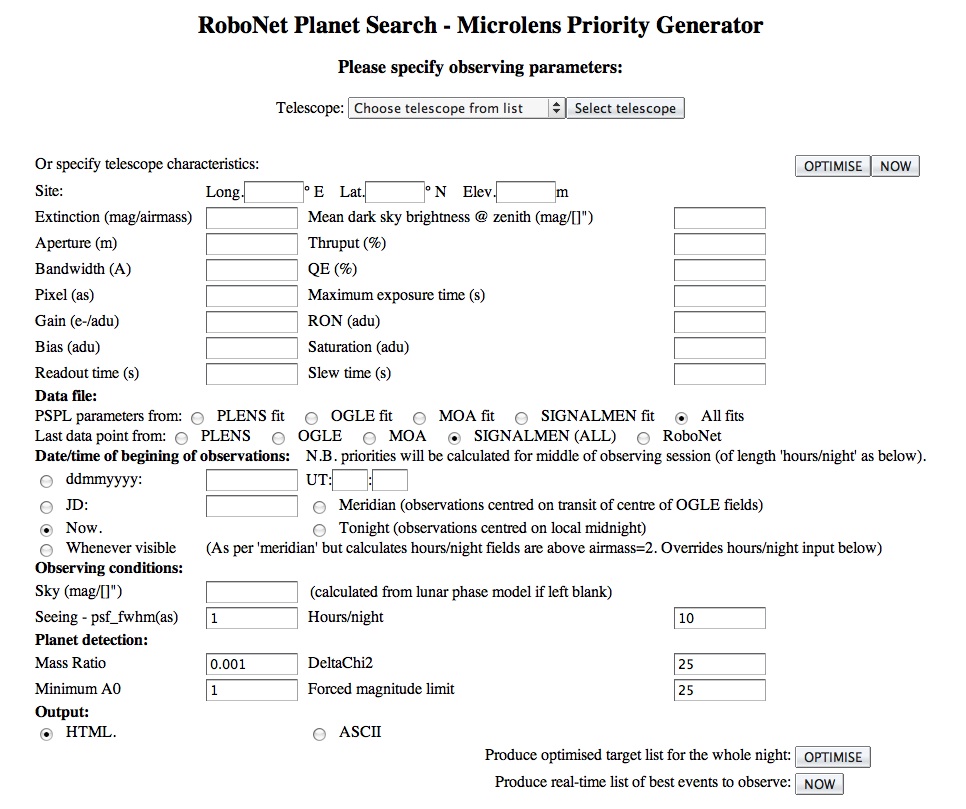} &
   \includegraphics[width=0.5\textwidth]{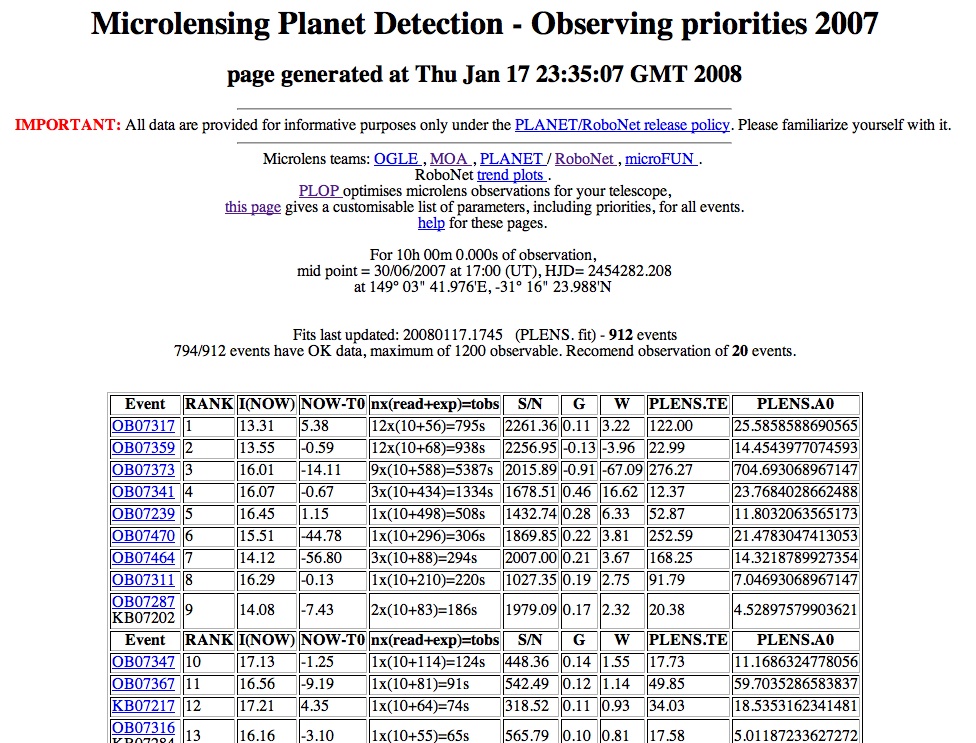} \\
\end{tabular}
   \caption{The input (choose-telescope.cgi) and output (optimise.cgi) screens of the web-PLOP system}
   \label{screens}
\end{figure}

There are three inputs to the web-PLOP system: 1.~Photometry from OGLE \& MOA (and other teams), which allow us to fit PSPL models to the events. 2.~Information on the telescope, sky conditions and the desired planet search conditions (mass ratio to optimise for and $\Delta\chi^2$ detection threshold). 3.~The desired time of observation and available observing time. 
The first is provided by the background data subscriber, which checks the OGLE and MOA pages every half hour, or can now use the data subscription service\footnote{Both SIGNALMEN and this data subscription service are part of the ARTEMiS concept: see \cite{artemis1,artemis2} for details.} that feeds SIGNALMEN. Inputs 2 and 3 are given to web-PLOP via a web form user interface (fig.~\ref{screens}).
The input page gets telescope 
parameters, or can select default 
values for these for a number of 
known telescopes. One can also 
select the source of the PSPL 
model to use (e.g.~the fits done by the PLOP background codes, or by SIGNALMEN, or the fits from the OGLE/MOA pages), and enter the time 
of observation in various ways (either for a specific time, or having the system calculate the time that the Bulge is visible tonight for the specified site). Finally the user can select the sort of output required, both in terms of format (customisable HTML tables for humans or fixed format plain text for robots) and also between an optimal list for the whole period of observation or a real-time output of what to observe `now'.

\section{Output}

The output HTML 
page for a human observer is also shown in fig~2. The columns on the HTML pages are customisable. The main output page lists the optimal events in order, with the suggested exposure times and number of exposures, along with predictions for the amplification and 
brightness. All other event parameters (e.g.~PSPL fit parameters, target information, comments, status, time of last update etc.) can be 
displayed; the columns are selected on the input form. 
Each event on the main page has a link to 
a more detailed page giving light-curves, $\Delta\chi^2$ maps, a table (also customisable) 
with event parameters, and links to more 
plots, the original OGLE / MOA page(s), 
finder charts, etc. The same interface 
(event\_pages.cgi) can also be used to 
browse the parameters for all other events, 
prioritised or not, with customisable tables. 

\section{Real-time prioritisation}\label{realtime}

In the 2008 season an improved web-PLOP will offer 
a number of new features. There will be 
real-time updating of priorities based on feedback 
from RoboNet, including observations on suspected 
anomalous events. To identify these events PLOP will link with the SIGNALMEN anomaly detector. This has some overlap with the background side of 
web-PLOP (shown in red fig \ref{plop}), as SIGNALMEN also keeps an up-to-date archive of all 
public data and fits its own PSPL models.  

The real-time prioritisation will be achieved by constantly updated scheduling: Instead of producing an optimal list for the 
night, we will provide priorities for what to observe `now', based on the same underlying prioritisation algorithm, but taking into account the detection zone area due to previous data points to prioritise those observations that will increase the area the most (i.e.~those producing non-overlapping detection zones). The constantly updated telescope scheduling also allows anomalous events to be scheduled with the appropriate sampling rate and exposure times along side normal monitoring observations, and for the telescopes to react fast to new suspected anomalies. This fast reaction and anomaly follow up will allow us to both detect \emph{and} characterise planetary anomalies down to Earth mass. 

Finally, we will also take into account the results of the latest observations using direct feedback from RoboNet. The eSTAR agent that passes commands to 
the telescopes will report the time and the estimated accuracy (from sky conditions) 
of each data point it takes directly to web-PLOP, even before the pipeline can 
process the data, so that real-time adjustments to the priorities can be made based on the actual conditions instead of the predicted ones. 

\begin{spacing}{0.9}

\end{spacing}

\end{document}